\documentclass{ws-procs9x6}
\newcommand{\be}{\begin{equation}}
\newcommand{\ee}{\end{equation}}
\newcommand{\bea}{\begin{eqnarray}}
\newcommand{\eea}{\end{eqnarray}}
\newcommand{\ba}[1]{\begin{array}{#1}}
\newcommand{\ea}{\end{array}}
\newcommand{\vg}{\bm{\gamma}}

\begin{document}

\title{A General Effective Theory for Dense Quark 
Matter}

\author{P.~T. REUTER, Q. WANG and D.~H. RISCHKE}

\address{Institute f\"ur Theoretische Physik,\\
Johann Wolfgang Goethe -- Universit\"at, \\ 
Robert-Mayer-Str. 10,\\ 
D-60054 Frankfurt am Main, Germany}

\maketitle

\abstracts{A general effective action for quark matter at nonzero 
temperature and/or nonzero density is derived. Irrelevant quark modes
are distinguished from 
relevant quark modes, and hard from soft gluon modes, by introducing two
separate cut-offs in momentum space, one for quarks, $\Lambda_q$, and one 
for gluons, $\Lambda_g$. Irrelevant quark modes and hard gluon modes are
then exactly integrated out in the functional integral representation of 
the QCD partition function. Depending on the specific choice for $\Lambda_q$
and $\Lambda_g$, the resulting effective action contains well-known effective
actions for hot and/or dense quark matter, for instance the ``Hard Thermal
Loop'' (HTL) or the ``Hard Dense Loop'' (HDL) action, as well as the 
high-density effective theory proposed by Hong and others. 
}

\section{Introduction}

Sufficiently cold and dense quark matter is a color superconductor
\cite{bailinlove}. Due to asymptotic freedom, 
at asymptotically large quark chemical potential,
$\mu \gg \Lambda_{\rm QCD}$, the strong coupling constant
$g \ll 1$, and one can apply weak-coupling methods to compute
the color-superconducting gap parameter.
For $N_f$ massless quark flavors
which participate in the screening of gluon fields,
to subleading order in $g$
the result is \cite{son,schaferwilczek,rdpdhr,rockefeller,qwdhr}
\be \label{phi}
\phi = 512\,\pi^4 \left( \frac{2}{N_f \, g^2} \right)^{5/2} \mu \,
\exp\left( - \frac{3 \pi^2}{\sqrt{2} g} - \frac{\pi^2 + 4}{8}
\right) \, \left[ 1 + O(g) \right]\;.
\ee
Sub-subleading corrections enter as $O(g)$ corrections to
the prefactor of the exponential. For a reliable extrapolation of this result
to quark chemical potentials of physical interest, $\mu \sim 500$ MeV,
it is mandatory to know the order of magnitude
of these sub-subleading corrections. 
However, in full QCD their calculation appears
prohibitively difficult. In order to proceed, note that
in weak coupling, $g \ll 1$, there is an ordering of scales
$\phi \sim \mu \exp(-1/g) \ll g \mu \ll \mu$. 
This ordering of scales implies that the modes near the Fermi surface,
which participate in the formation of Cooper pairs and are therefore of
primary relevance in the gap equation, can be considered to be
independent of the detailed dynamics of the modes deep within the 
Fermi sea. This suggests that the most efficient way to compute 
properties such as the color-superconducting gap parameter is 
via an {\em effective theory for quark modes near the Fermi surface}.
Such an effective theory has been originally proposed by
Hong \cite{hong,hong2} and was subsequently refined by others
\cite{HLSLHDET,schaferefftheory,NFL,others}. 

In this paper, we derive a general effective theory for hot and/or
dense quark matter from first principles (see also Ref.\ \cite{ptrqwdhr}).
We introduce cut-offs in momentum space
for quarks, $\Lambda_q$, and gluons, $\Lambda_g$. These cut-offs
separate relevant from irrelevant quark modes and soft from hard
gluon modes. We then explicitly integrate out irrelevant quark and
hard gluon modes. 
We show that the standard HTL and HDL effective actions are
contained in our general effective action for 
a certain choice of the quark and gluon
cut-offs $\Lambda_q ,\, \Lambda_g$.
It can also be shown \cite{ptrqwdhr} 
that the action of the high-density effective theory
derived by Hong and others 
\cite{hong,hong2,HLSLHDET,schaferefftheory,NFL,others}
is a special case of our general effective action.
We also outline the computation of the color-superconducting
gap parameter with this effective action (for more details,
see Ref.\ \cite{ptrqwdhr}).

Our units are $\hbar=c=k_B=1$. 4-vectors are denoted by
$K^\mu = (k_0, {\bf k})$, with ${\bf k}$ being a
3-vector of modulus $|{\bf k}| \equiv k$. For the summation over Lorentz
indices, we use the familiar Minkowski metric
$g^{\mu \nu} = {\rm diag}(+,-,-,-)$, although 
we exclusively work in compact Euclidean space-time with 
volume $V/T$, where $V$
is the 3-volume and $T$ the temperature of the system. 
Space-time integrals 
are denoted as $\int_0^{1/T} d \tau \int_V d^3{\bf x} \equiv
\int_X$. Energy-momentum sums are denoted as
$(T/V)\sum_{K} \equiv T\sum_n (1/V) \sum_{\bf k}$.  
The sum over $n$ runs over the
Matsubara frequencies, $\omega_n^{\rm b} = 2n \pi T$ for bosons and 
$\omega_n^{\rm f} = (2 n+1)\pi T$ for fermions. 
The 4-dimensional delta-function
is conveniently defined as $\delta^{(4)}(X) \equiv \delta(\tau)\,
\delta^{(3)}({\bf x})$.

\section{Deriving the Effective Theory}

The partition function for QCD in the absence of external
sources reads
\be \label{ZQCD}
Z = \int  [D A] \, 
\exp \left\{ S_A [A]\right\}\, Z_q[A] \,\, .
\ee
Here the (gauge-fixed) gluon action is
\be \label{SA}
S_A[A] = \int_X \left[ - \frac{1}{4} F^{\mu \nu}_a (X) \, F_{\mu \nu}^a
(X) \right] + S_{\rm gf}[A] + S_{\rm ghost}[A] \,\,,
\ee
where $F_{\mu \nu}^a = \partial_\mu A_\nu^a - \partial_\nu A_\mu^a
+ g f^{abc} A_\mu^b A_\nu^c$ is the gluon field strength tensor,
$S_{\rm gf}$ is the gauge-fixing part, and $S_{\rm ghost}$ the
ghost part of the action.

The partition function for quarks in the presence of gluon fields is
\be \label{Zq}
Z_q[A] = \int [D \bar{\Psi}] \, [ D \Psi]\,
\exp \left\{ S_q[A,\bar{\Psi},\Psi] \right\}\,\, ,
\ee
where the quark action is
\be \label{quarkaction}
S_q[A, \bar{\Psi}, \Psi] = 
\frac{1}{2} \int_{X,Y} \bar{\Psi}(X) \,  G_0^{-1} (X,Y)\,
\Psi(Y) + \frac{g}{2} \int_X \bar{\Psi}(X) \, \hat{\Gamma}^\mu_a
A_\mu^a(X) \, \Psi(X) \,\, .
\ee
Here, 
\be
\Psi \equiv \left( \begin{array}{c}
                    \psi \\
                    \psi_C \end{array} \right) \;\; , \;\;\;\;
\bar{\Psi} \equiv ( \bar{\psi} , \bar{\psi}_C )\,\, ,
\ee
are Nambu-Gor'kov quark spinors,
where $\psi_C \equiv C \bar{\psi}^T,\; \bar{\psi}_C \equiv \psi^T  C$
are charge-conjugate quark spinors;
$C \equiv i \gamma^2 \gamma_0$ is the charge conjugation matrix.
The free inverse quark propagator in the Nambu-Gor'kov basis is
\be
G_0^{-1}(X,Y)  \equiv \left( \begin{array}{cc} 
                          [G_0^+]^{-1}(X,Y) & 0 \\
                           0 & [G_0^-]^{-1}(X,Y) \end{array} \right)\,\, ,
\ee
with the free inverse propagator for quarks and charge-conjugate
quarks
\be
[G_0^\pm]^{-1}(X,Y) \equiv (i \slash \!\!\!\partial_X \pm \mu
\gamma_0 - m )\, \delta^{(4)}(X-Y)\,\, .
\ee
The quark-gluon vertex in the Nambu-Gor'kov basis is defined as
\be \label{NGvertex}
\hat{\Gamma}^\mu_a \equiv \left( \begin{array}{cc}
                               \gamma^\mu T_a & 0 \\
                               0 & -\gamma^\mu T_a^T 
                              \end{array} \right) \,\, .
\ee
We now transform all fields into energy-momentum space. 
The partition function for quarks becomes
\be \label{Zq2}
Z_q [A]= \int [D \bar{\Psi}]\, 
[ D \Psi]\; \exp \left[ \frac{1}{2} \, \bar{\Psi} \left( 
G_0^{-1} + g  A \right) \Psi \right]
\,\, .
\ee
Here, we employ a compact matrix notation,
\be \label{compact}
\bar{\Psi} \, \left(  G_0^{-1} + g A \right)\,   \Psi \equiv
\sum_{K,Q} \bar{\Psi}(K) \, \left[  G_0^{-1}(K,Q) 
+ g A(K,Q) \right] \, \Psi(Q)\;\; ,
\ee
with 
\be \label{G_0FT}
G_0^{-1}(K,Q) = \frac{1}{T} \left( \begin{array}{cc}
                             [G_0^+]^{-1}(K) & 0 \\
                              0 & [G_0^-]^{-1}(K) \end{array} \right)
\delta^{(4)}_{K,Q} \,\, ,
\ee
where $[G_0^\pm]^{-1}(K) \equiv \slash \!\!\!\! K \pm \mu \gamma_0 - m$,
and
\be \label{calA}
A(K,Q) \equiv \frac{1}{\sqrt{VT^3}}\, \hat{\Gamma}^\mu_a
A_\mu^a(K-Q) \,\, .
\ee
The next step is to integrate out irrelevant quark modes.
We define projection operators $P_1,\, P_2$ for 
relevant and irrelevant quark modes, respectively,
\be \label{project}
\Psi_1 \equiv  P_1 \, \Psi\;\; , \;\;\;\;
\Psi_2 \equiv  P_2 \, \Psi \;\; , \;\;\;\;
\bar{\Psi}_1 \equiv \bar{\Psi} \, \gamma_0  P_1 \gamma_0 
\;\; , \;\;\;\;
\bar{\Psi}_2 \equiv \bar{\Psi} \, \gamma_0  P_2 \gamma_0 
\;\;.
\ee
The Grassmann integration over the irrelevant quark fields $\bar{\Psi}_2,\,
\Psi_2$ can be done exactly. The result for the QCD partition function
is
\begin{subequations}
\bea \label{Z4}
Z & = &\int [D A] \, [ D \bar{\Psi}_1]\, 
[ D \Psi_1] \, \exp \left\{ S[A, \bar{\Psi}_1, \Psi_1] \right\} \;,\\
\!\!\!\!\!\!
S[A, \bar{\Psi}_1, \Psi_1] & \equiv & S_A[A] + \frac{1}{2} \bar{\Psi}_1 
\left( G^{-1}_{0,11} + g B[A] \right) \Psi_1
+ \frac{1}{2} {\rm Tr}_q \ln G_{22}^{-1}[A], 
\label{S}
\eea
\end{subequations}
where
\begin{subequations}
\bea \label{B}
g B [A] & \equiv & g A_{11} - g A_{12} \, G_{22}[A] \, g A_{21}\; ,\\
G^{-1}_{22}[A] & \equiv&  G_{0,22}^{-1} + g A_{22} \label{G22}\;.
\eea
\end{subequations}
The quantities $G^{-1}_{0,11},\,G^{-1}_{0,22},\, A_{11},\, A_{12},\, A_{21}, \,
A_{22}$ carry as argument a pair of quark 4-momenta, $(K,Q)$.
The set of subscripts $(nm), n,m = 1, 2,$ at these quantities
indicates whether the corresponding 3-momenta ${\bf k}, \, {\bf q}$ 
belong to the subspace of relevant ($n=1$) or irrelevant ($n=2$) quark modes.

Similarly to the treatment of fermions 
we now define projectors $Q_1,\, Q_2$ for
soft and hard gluon modes, respectively,
\be
A_1  \equiv Q_1 \, A \;\; , \;\;\;\; A_2 \equiv Q_2 \, A\;.
\ee
Inserting $A \equiv A_1 + A_2$ into Eq.\ (\ref{S}),
we sort the result with respect to powers of the hard gluon field,
$A_2$,
\bea \label{expansion}
S[A,\bar{\Psi}_1,\Psi_1] & = & S[A_1,\bar{\Psi}_1,\Psi_1]
+ A_2 \,J [A_1,\bar{\Psi}_1,\Psi_1] \nonumber \\
& - & \frac{1}{2}\,
A_2 \, \Delta^{-1}_{22}[A_1, \bar{\Psi}_1,\Psi_1]\, A_2 
+ S_I[A_1,A_2,\bar{\Psi}_1,\Psi_1] \;.
\eea
The first term in this expansion, containing no hard gluon fields
at all, is simply the action (\ref{S}), with $A$ replaced by
the soft gluon field $A_1$.
The second term, $A_2  J$,
contains a single power of the hard gluon field, where
$J [A_1,\bar{\Psi}_1,\Psi_1] 
\equiv J_{B}[A_1, \bar{\Psi}_1, \Psi_1]
+ J_{\rm loop}[A_1] + J_{V}[A_1]$.
The first contribution, $J_B$,
arises from the coupling of the relevant fermions to the
``modified'' gluon field $B[A]$, i.e., from the second term in 
Eq.\ (\ref{S}).
The second contribution, $J_{\rm loop}$, 
arises from the term ${\rm Tr}_q \ln G_{22}^{-1}$
in Eq.\ (\ref{S}), plus an analogous contribution from the Fadeev-Popov
determinant contained in $S_A[A]$. Finally, the third contribution, $J_{V}$, 
arises from the non-Abelian vertices in QCD.

The third term in Eq.\ (\ref{expansion}) is quadratic in $A_2$, 
where $\Delta_{22}^{-1}[A_1,\bar{\Psi}_1,\Psi_1]\equiv 
\Delta_{0,22}^{-1} + \Pi_{22}[A_1, \bar{\Psi}_1, \Psi_1]$.
Here, $\Delta_{0,22}^{-1}$ is the free inverse propagator for hard gluons.
Similarly to the ``current'' $J$
the ``self-energy'' $\Pi_{22}$ of hard gluons consists of three
different contributions,
$\Pi_{22}[A_1, \bar{\Psi}_1, \Psi_1] =
\Pi_{B}[A_1, \bar{\Psi}_1, \Psi_1] +
\Pi_{\rm loop}[A_1] +
\Pi_{V}[A_1]$.

\begin{figure}[ht]
\centerline{\epsfxsize=1in\epsfbox{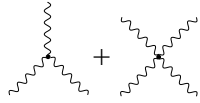}}   
\caption{The three- and four-gluon vertices in $S_A[A_1]$. \label{fig1}}
\end{figure}

\begin{figure}[ht]
\centerline{\epsfxsize=4in\epsfbox{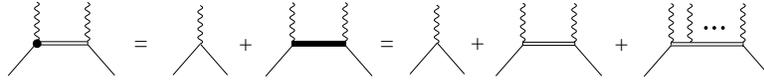}}   
\caption{The term $\bar{\Psi}_1 g B[A_1] \Psi_1$ in the effective
action (\ref{Seff}). The left-hand side is a graphical definition
for this term. The right-hand side is obtained from the
definition of $gB[A_1]$, Eq.\ (\ref{B}), and subsequently expanding the
full propagator for irrelevant quarks, $G_{22}$ (thick solid line),
in this equation in powers of the soft gluon field, $A_1$ (wavy line),
according to Eq.\ (\ref{G22}). A double line
stands for the free propagator for irrelevant quarks, $G_{0,22}$. 
Thin lines stand for relevant quark fields $\Psi_1, \bar{\Psi}_1$.
\label{fig2}}
\end{figure}

\begin{figure}[ht]
\centerline{\epsfxsize=4in\epsfbox{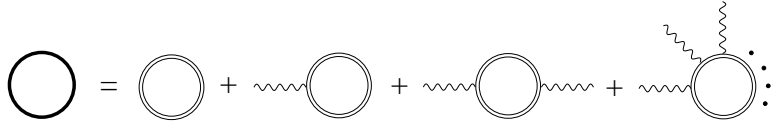}}   
\caption{The term ${\rm Tr}_q \ln G_{22}^{-1}[A_1]$ in
the effective action (\ref{Seff}). The right-hand side
is obtained upon expanding the full propagator $G_{22}$ in
powers of the soft gluon field, $A_1$, according to
Eq.\ (\ref{G22}). \label{fig3}}
\end{figure}

\begin{figure}[ht]
\centerline{\epsfxsize=4in\epsfbox{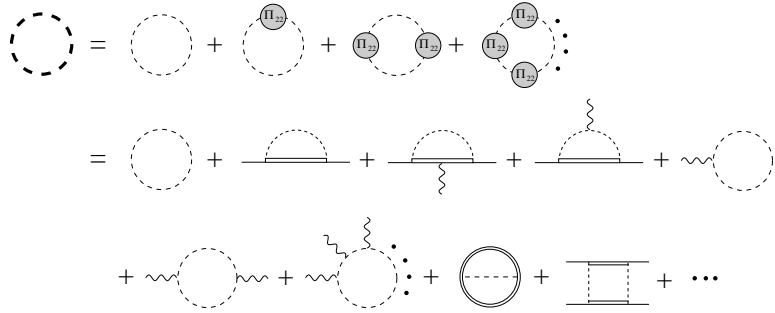}}   
\caption{The term ${\rm Tr}_g \ln
\Delta_{22}^{-1}[A_1,\bar{\Psi}_1,\Psi_1]$
in the effective action (\ref{Seff}). The first line is obtained
from the expression for $\Delta_{22}$; a dashed line denotes the
free propagator for hard gluons $\Delta_{0,22}$, a bubble
denotes the full self-energy $\Pi_{22}$. The second and third
lines contain some examples for diagrams generated
when explicitly inserting $\Pi_{22}$. \label{fig4}}
\end{figure}

\begin{figure}[ht]
\centerline{\epsfxsize=4in\epsfbox{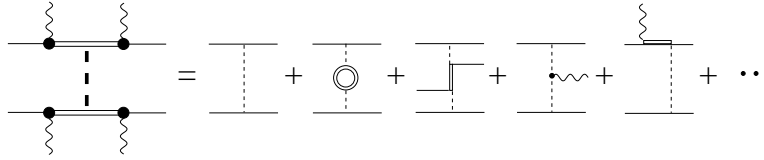}}   
\caption{The term $J_B \Delta_{22} J_B$ in the effective
action (\ref{Seff}). The left-hand side serves
as a graphical definition for this term, the right-hand side
shows some examples for diagrams when explicitly inserting
the expressions for $J_B$ and $\Delta_{22}$. \label{fig5}}
\end{figure}

Finally, we collect all terms with more than
two hard gluon fields $A_2$ in Eq.\ (\ref{expansion}) in the
``interaction action'' for hard gluons, 
$S_I[A_1,A_2,\bar{\Psi}_1,\Psi_1]$.
In order to perform
the functional integration over the hard gluon fields $A_2$
one adds the source term $A_2 J_2$ to the action (\ref{S})
and then replaces the fields $A_2$ in $S_I$
by functional differentiation with respect to $J_2$, at
$J_2=0$. Moving
the factor $\exp\{ S_I[A_1, \delta/\delta J_2, \bar{\Psi}_1, \Psi_1 ]\}$
in front of the functional $A_2$-integral, the latter
is Gaussian and can be readily
performed (after a suitable shift of $A_2$). We
refrain from giving the explicit result, because we will
simplify it anyway by employing the following two approximations.
The first is based on the
principle assumption in the construction of any
effective theory, namely that soft and hard modes are well separated 
in momentum space. Consequently, momentum conservation
does not allow a hard gluon to
couple to any (finite) number of soft gluons. Under this assumption, the
diagrams generated by $A_2 ( J_{\rm loop} + J_{V})$
will not occur in the effective theory.
In the following, we shall therefore omit these terms, such that
$ J \equiv J_{B}$.
Our second approximation is that we approximate $S_I \simeq 0$. 
With these approximations, the partition function reads
\be \label{Z5}
Z =  \int [D A_1]\, [D \bar{\Psi}_1] \, [D \Psi_1]\, 
\exp\{S_{\rm eff} [A_1,\bar{\Psi}_1, \Psi_1 ] \}\;,
\ee
with the effective action
\bea
S_{\rm eff} [A_1,\bar{\Psi}_1, \Psi_1 ] & \equiv & 
S_A[A_1] + \frac{1}{2} \, \bar{\Psi}_1 
\left\{ G_{0,11}^{-1} +  g B[A_1] \right\} \Psi_1 \nonumber \\
& +  &  \frac{1}{2}\, {\rm Tr}_q \ln  G_{22}^{-1}[A_1] 
- \frac{1}{2}\,  {\rm Tr}_g \ln
\Delta_{22}^{-1}[A_1,\bar{\Psi}_1,\Psi_1] 
\nonumber \\
& + &  \frac{1}{2} \,  J_{B}[A_1,\bar{\Psi}_1,\Psi_1]   \,
\Delta_{22}[A_1,\bar{\Psi}_1,\Psi_1] 
\,  J_{B}[A_1,\bar{\Psi}_1,\Psi_1].
\label{Seff}
\eea
This is the desired action for the effective theory describing the
interaction of relevant quark modes, $\bar{\Psi}_1, \Psi_1$, and
soft gluons, $A_1$. 
The vertices arising in the effective action (\ref{Seff}) are
displayed in Figs.\ \ref{fig1} -- \ref{fig5}.

\section{Examples}

In all examples discussed in this section, the gluon projectors
are chosen as
\begin{subequations} \label{Q12}
\bea
 Q_1(P_1,P_2) & \equiv & \Theta(\Lambda_g -p_1) \, \delta^{(4)}_{P_1,P_2}
\; , \\
 Q_2(P_1,P_2) & \equiv & \Theta(p_1-\Lambda_g) \, \delta^{(4)}_{P_1,P_2}
\; .
\eea
\end{subequations}
The gluon cut-off momentum $\Lambda_g$ separates
soft gluon fields, $A_1 \equiv Q_1 A$, from hard gluon
fields, $A_2 \equiv Q_2 A$.

\subsection{HTL/HDL Effective Action}

This action defines an effective theory for massless
quarks and gluons with small momenta in a system
at high temperature $T$ (HTL), or large chemical potential $\mu$ (HDL). 
Consequently, the projectors $ P_{1,2}$ for quarks are given by
\begin{subequations} \label{PHTL}
\bea
 P_1 (K,Q) & = & \Theta(\Lambda_q -k)\,\delta^{(4)}_{K,Q}\;, \\
 P_2 (K,Q) & = & \Theta(k - \Lambda_q)\,\delta^{(4)}_{K,Q}\;,
\eea
\end{subequations}
The essential assumption to derive the HTL/HDL effective action is
that there is a single momentum scale, 
$\Lambda_q = \Lambda_g \equiv \Lambda$,
which separates hard modes with momenta $\sim T$, or $\sim
\mu$, from soft modes with momenta $\sim gT$, or $\sim g\mu$.
In order to show that the HTL/HDL effective action is contained in the
effective action (\ref{Seff}), we first note that a soft particle
cannot become hard by interacting with another soft particle.
This has the consequence that a soft quark cannot 
turn into a hard one by soft-gluon scattering. Therefore,
\be
g B[A_1] \equiv g A_{11}\;.
\ee
Another consequence is that the last term in Eq.\ (\ref{Seff}),
$J_{B} \Delta_{22} J_{B}$, vanishes since $J_{B}$
is identical to a vertex between a soft quark and a hard gluon, which
is kinematically forbidden. The resulting action then reads
\bea
S_{\mbox{\scriptsize high}\,T/\mu} 
[A_1,\bar{\Psi}_1, \Psi_1 ] & \equiv &
S_A[A_1] + \frac{1}{2} \, \bar{\Psi}_1 
\left( G_{0,11}^{-1} +  g A_{11} \right) \Psi_1 \nonumber \\
& + & \frac{1}{2} {\rm Tr}_q \ln  G_{22}^{-1}[A_1] 
- \frac{1}{2}  {\rm Tr}_g \ln
\Delta_{22}^{-1}[A_1,\bar{\Psi}_1,\Psi_1] . 
\label{SHTL}
\eea
We realize that the 
term ${\rm Tr}_q \ln G_{22}^{-1}$ generates
all one-loop diagrams, where $n$ soft gluon legs 
are attached to a hard quark loop. This is precisely the quark-loop 
contribution to the HTL/HDL effective action.

Now consider the last term, ${\rm Tr}_g \ln
\Delta_{22}^{-1}$.
For hard gluons with momentum $\sim T$ or $\sim \mu$,
the free inverse gluon propagator is $\Delta_{0,22}^{-1} \sim
T^2$ or $\sim \mu^2$, while the contribution $\Pi_{\rm loop}$ to the hard 
gluon ``self-energy'' $\Pi_{22}$ is at most of the order $\sim g^2
T^2$ or $\sim g^2 \mu^2$. Consequently, $\Pi_{\rm loop}$ can be neglected and
$\Pi_{22}$ only contains tree-level diagrams, $\Pi_{22} \equiv \Pi_{B}
+ \Pi_{V}$. Now expand ${\rm Tr}_g \ln
\Delta_{22}^{-1}$ in powers of $\Pi_{22}$.
The terms which contain only insertions of $\Pi_{V}$ correspond to one-loop
diagrams where $n$ soft gluon legs
are attached to a hard gluon loop. This is precisely  
the pure gluon loop contribution to the HTL effective action. 
For the HDL effective action, this contribution is  $\sim g^2 T^2$,
and thus negligible.

The ``self-energy'' $\Pi_{B}$ contains only two soft
quark legs attached to a hard quark propagator (via emission and
absorption of hard gluons). Consequently, in the expansion of
${\rm Tr}_g \ln \Delta_{22}^{-1}$, 
the terms which contain insertions of $\Pi_{V}$ and $\Pi_{B}$
correspond to one-loop diagrams where an arbitrary number of 
soft quark and gluon legs is attached to the loop. It was shown in Ref.\
\cite{braatenpisarski} that of these diagrams, only the ones with
two soft quark legs and no four-gluon vertices are kinematically
important and thus contribute to the HTL/HDL effective action.
We have thus shown that the latter
is contained in the effective action (\ref{SHTL}), and constitutes its
leading contribution.

\subsection{High-density Effective Theory}

\begin{figure}[ht]
\centerline{\epsfxsize=2in\epsfbox{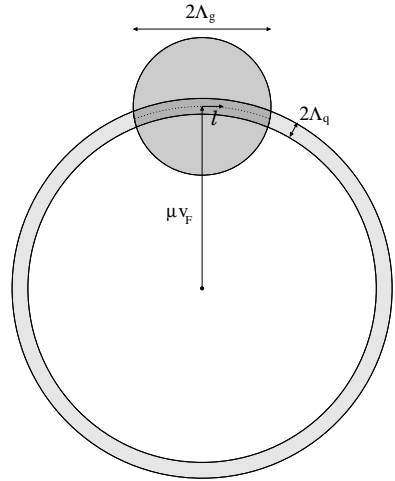}}   
\caption{The shell of relevant quark modes of width $2 \Lambda_q$ 
around the Fermi surface, and the patch on this shell,
characterizing the range of soft gluon interactions. \label{fig6}}
\end{figure}

In an effective theory 
for cold, dense quark matter the projectors are chosen as
\begin{subequations} \label{P12}
\bea
P_1(K,Q) & \equiv & \left( \begin{array}{cc}
 \Lambda_{\bf k}^+ & 0 \\
 0 & \Lambda_{\bf k}^- \end{array} \right)  
\Theta(\Lambda_q - | k - k_F|) \, \delta^{(4)}_{K,Q} \;, \\
P_2(K,Q) & \equiv & \left[ \left( \begin{array}{cc} 
\Lambda_{\bf k}^- & 0 \\
0 & \Lambda_{\bf k}^+ \end{array} \right)
+ \left( \begin{array}{cc}
\Lambda_{\bf k}^+\, & 0 \\
0 &   \Lambda_{\bf k}^- \end{array} \right) \Theta(| k - k_F| -
\Lambda_q) \right] \, \delta^{(4)}_{K,Q}\;.
\eea
\end{subequations}
Here, 
$\Lambda^e_{\bf k} \equiv  
\left[ E_{\bf k} + e \gamma_0 \left(\vg \cdot
{\bf k} + m \right) \right]/(2 E_{\bf k})$
are projection operators onto states with positive ($e = +$)
or negative ($e=-$) energy, where $E_{\bf k} = \sqrt{{\bf k}^2 +m^2}$
is the relativistic single-particle energy.
The momentum cut-off $\Lambda_q$ 
controls how many quark modes (with positive
energy) are integrated out. Thus, all quark modes within a layer of 
width $2 \Lambda_q$ around the Fermi surface are considered as 
relevant, while all antiquark modes and quark modes
outside this layer are considered as irrelevant, cf.\ Fig.\ \ref{fig6}.
It can be shown \cite{ptrqwdhr} that the effective theory with
the projection operators (\ref{P12}) is equivalent to the
high-density effective theory proposed by Hong \cite{hong,hong2}
and refined by others \cite{HLSLHDET,schaferefftheory,NFL,others}.

Armed with the effective action (\ref{Seff}), one can
compute the color-superconducting gap parameter in
weak coupling \cite{ptrqwdhr}. One first derives the 2PI effective
action \cite{CJT}, $\Gamma$, with Eq.\ (\ref{Seff}) as the underlying
{\it tree-level\/} action. The stationary points of
$\Gamma$ yield Dyson-Schwinger equations for the
propagators of relevant quarks and soft gluons, which can be solved
in a particular many-body approximation scheme. The 
gap equation for the color-superconducting gap parameter
is equivalent to the equation
for the off-diagonal component of the quark propagator
in the Nambu-Gor'kov basis. The introduction of
the quark and gluon cut-offs $\Lambda_q$ and $\Lambda_g$
allows for a rigorous power-counting of various contributions
to the gap equation. It turns out that, in order to obtain
the subleading-order result (\ref{phi}), one has
to assume $\Lambda_q \lesssim g \mu \ll \Lambda_g \lesssim \mu$.

The situation is illustrated in Fig.\ \ref{fig6}.
Relevant quark modes lie in a narrow ``shell'' of width $2 \Lambda_q$
around the Fermi surface, irrelevant quarks lie outside
this shell. The leading contribution to the QCD gap equation
arises from soft magnetic gluon exchange, which mediates between quark states
inside a ``patch'' of diameter $2 \Lambda_g$ located on this shell.
Subleading contributions arise from soft electric gluon exchange,
as well as from hard magnetic and electric gluon exchange.
The latter mediates between quark states
in- and outside this patch, inside the shell of relevant
quark modes.

\section*{Acknowledgments}
The work of Q.W.\ is supported by the
Virtual Institute VH-VI-041 of the
Helmholtz Association.

\end{document}